\documentclass{sigchi}

\toappear{\phantom{Permission to make digital or hard copies of all or part of this work for personal or classroom use is granted without fee provided that copies are not made or distributed for profit or commercial advantage and that copies bear this notice and the full citation on the first page. Copyrights for components of this work owned by others than ACM must be honored. Abstracting with credit is permitted. To copy otherwise, or republish, to post on servers or to redistribute to lists, requires prior specific permission and/or a fee. Request permissions from permissions@acm.org. \\
Conference data.\\
Copyright string.\\
DOI string.}}

\usepackage{balance}  
\usepackage{graphics} 
\usepackage{times}    
\usepackage{url}      
\usepackage{array}
\usepackage [english]{babel}
\usepackage [autostyle, english = american]{csquotes}
\MakeOuterQuote{"}

\makeatletter
\def\url@leostyle{%
  \@ifundefined{selectfont}{\def\UrlFont{\sf}}{\def\UrlFont{\small\bf\ttfamily}}}
\makeatother
\def\pprw{8.5in}
\def\pprh{11in}

\usepackage[pdftex]{hyperref}
\hypersetup{
pdftitle={Interactive Modeling of Concept Drift and Errors in Relevance Feedback},
pdfauthor={AK, YC, DG, SK},
pdfkeywords={Concept drift, Exploratory search, Interactive User Modeling, Probabilistic User Models, User interfaces},
bookmarksnumbered,
pdfstartview={FitH},
colorlinks,
citecolor=black,
filecolor=black,
linkcolor=black,
urlcolor=black,
breaklinks=true,
}

\newcommand\tabhead[1]{\small\textbf{#1}}

\begin{document}

\title{Interactive Modeling of Concept Drift and Errors in Relevance Feedback}

\numberofauthors{4}
\author{
\hspace*{-3em}Antti Kangasr{\"a}{\"a}si{\"o}{\Large$^\circ$}\; Yi Chen{\Large$^\circ$}\; 
 Dorota G\l owacka{\Large$^\star$}\; Samuel Kaski{\Large$^\circ$}\hspace*{-3em}
\and
\affaddr{{${^\circ}$}Helsinki Institute for Information Technology HIIT, Department of Computer Science, Aalto University}
\and
\affaddr{{${^\star}$} Helsinki Institute for Information Technology HIIT, Department of Computer Science,}\\
\affaddr{University of Helsinki}\\
\email{ first.last@hiit.fi}
}

\maketitle

\begin{abstract}
Users giving relevance feedback in exploratory search are often uncertain about the correctness of their feedback, which may result in noisy or even erroneous feedback. Additionally, the search intent of the user may be volatile as the user is constantly learning and reformulating her search hypotheses during the search. This may lead to a noticeable concept drift in the feedback. We formulate a Bayesian regression model for predicting the accuracy of each individual user feedback and thus find outliers in the feedback data set. Additionally, we introduce a timeline interface that visualizes the feedback history to the user and gives her suggestions on which past feedback is likely in need of adjustment. This interface also allows the user to adjust the feedback accuracy inferences made by the model. Simulation experiments demonstrate that the performance of the new user model outperforms a simpler baseline and that the performance approaches that of an oracle, given a small amount of additional user interaction. A user study shows that the proposed modelling technique, combined with the timeline interface, makes it easier for the users to notice and correct mistakes in their feedback, and to discover new items.
\end{abstract}

\keywords{
	Concept drift;
    Exploratory search;
    Interactive User Modeling;
	Probabilistic User Models;
	User interfaces
}

\category{H.3.3}{Information Search and Retrieval}{Relevance feedback}
\category{H.5.2}{Information Interfaces and Presentation}{User Interfaces}


\section{Introduction}
Search can be broadly divided into two categories:
 (1) exploratory search, where the goals are ill-defined and may change as search progresses, 
and (2) lookup search, where the user has a specific target in mind \cite{marchionini06,white:2006}. 
A lookup search begins with the users expressing their information need as precisely as possible to reach the correct area of the information space. 
By contrast, user behavior in exploratory search is highly dynamic. 
Users begin exploration with no clear search goals in mind and issue search queries that are imprecise at first. 
They browse through the search results and iteratively reformulate their queries using new keywords they discover \cite{marchionini06}. 
Additionally, in exploratory tasks users are uncertain how to formulate their search queries \cite{chowdhury2011uncertainty}. 

A recently developed search system called  \textit{SciNet} \cite{Glowacka2013,Kangasraasio2015} aims to assist the user in exploratory search tasks by allowing her to interactively refine her search query. The user starts the search with a general keyword query and gradually refines the system's user model through interactive relevance feedback to keywords suggested by the system. However, the user model in this system makes the assumptions that: (1) all the user feedback is equally accurate, (2) the user makes no mistakes in giving feedback, and (3) no learning or changes in the user's search interests would occur as the search progresses. In short, the system does not take the possibility of concept drift \cite{gama2014survey,tsymbal2004problem} into account. In this paper, we improve over the existing system by formulating a user model that is able to deal with concept drift and we develop an interface that allows the user to interact with this model.

The proposed user model is a Bayesian regression model that is able to estimate both the current search intent of the user and the accuracy of the relevance feedback provided by the user. For the purpose of collecting user feedback, we introduce a timeline interface that allows the user to see her recent feedback history. The interface highlights the past feedback values that were estimated to be inaccurate and allows the user to interact with the visualized keywords. We demonstrate that the proposed  model is able to improve retrieval accuracy in a simulation experiment. In a user study, participants report that the new interface makes it easier for the users to notice and correct mistakes in their feedback. In addition, users report that the new system helped them to discover new items and that the results were more diverse with the new system.

The paper is organised as follows. In the next section we present a brief overview of related literature. Next, we describe the proposed user model and the new interface that allows the user to adjust the user model. Finally, we show results from simulations and user studies.

\newpage
\section{Related Work}
In most interactive systems, the user has a concept in mind that the system is trying to learn while the user interacts with the system, for example, a particular genre of music or specific types of documents. Many of such systems rely on machine learning techniques to help the system to identify the concept that the user has in mind. However, as human interests are often quite complex, it is common that the predictions will have errors, especially if the data is noisy or there is only little of it. Therefore, in recent years there has been a growing interests in developing new applications that would allow the user to correct the model of the user's needs that a machine learning system has built \cite{fails2003interactive, kapoor2010interactive, kulesza2015principles, stumpf2009interacting}. This type of a system explains the reasons for its predictions to the user, who in turn explains corrections back to the system. This both helps the system to make a better model of the user's interests and helps the user to build a mental model to predict how the system will behave. Researchers have explored using this cycle of quick interactions to train instance-based classifiers \cite{fogarty2008cueflik}, elicit labels for the most important instances \cite{cakmak2010designing}, and to improve reinforcement learning for automated agents \cite{knox2012reinforcement}. However, none of the above applications deal with the idea of involving the user in the interactive search loop in the concept drift setting.

Open user models are another important branch of research in the area of user modelling \cite{Ahn2009, Bakalov2013}. A user model is an internal representation of the user's knowledge or interests that an adaptive information retrieval system (IR) can use to recommend new items to a given user. In most IR systems, the user model is hidden from the user, however, adaptive IR systems with open user models allow the user to view the system's representation of her interests or search goals and edit it. Open user modelling has been very popular in the e-learning community \cite{Bull2004, Dimitrova2001} and has also been applied in other domains, such as news recommendation \cite{Ahn2007} or Wikipedia page recommendation \cite{Lehmann2010}. Recent studies show that interactive open user models can greatly improve user performance and user satisfaction \cite{Bakalov2013, Glowacka2013, Kangasraasio2015}. However, these systems assume that the user interests are fixed and do not change over a search session. Our modeling technique combined with the proposed interactive user model visualisation takes into account the gradual concept drift that frequently occurs in exploratory search.

\section{The User Model}

We start by assuming that we have a large collection of items that we could recommend to the user. Each item $i$ has a feature vector $\boldsymbol{x_i}$ and our main goal is to estimate the relevance $y_i$ of each item, based on observations made about the user's search interests. These observations in general are based on relevance feedback provided by the user: the user indicates that the relevance of item $i$ is $y_i$.

We make the simplifying assumption that the function that predicts the  relevance of an item  based on this item's features is approximately linear. We assume that the errors made by the model are normally distributed so  the general accuracy of this model is described by the variance $\sigma^2$. To accommodate  observations that have different accuracies, we assume that each observation has a weight factor $w_i$ that scales the global model variance. This gives us the following observation model:
\begin{align}
y_i \sim Normal(\boldsymbol{x_i} \boldsymbol{\phi}, \sigma^2 / w_i),
\label{eq:model}
\end{align}
where $\boldsymbol{\phi}$ are the linear model coefficients.

To make a fully Bayesian model, we assume prior distributions for the parameters:
\begin{align}
\phi_j \sim Normal(\mu_\phi, \lambda_\phi),\\
\sigma^2 \sim InverseGamma(\alpha_{\sigma^2}, \beta_{\sigma^2}),\\
w_i \sim Gamma(\alpha_{w}, \beta_{w}),\\
w_i^{fix} \sim Delta(1.0),
\label{eq:priors}
\end{align}
where $\phi_j$ is the $j$th component of the vector $\boldsymbol{\phi}$. Generally, we assume that the accuracies of the observations are unknown and drawn from a Gamma distribution. However, we also allow the user to inform the system about the accuracy of her feedback: if the user has explicitly marked certain feedback as accurate, we use $w_i^{fix}$ instead of $w_i$, making the accuracy for that feedback  equal to $1.0$. We also assume that the most recent feedback is always similarly accurate.
In this paper we will refer to this model as the \textit{ARD model}, as the determination of observation weights can be seen as Automatic Relevance Determination \cite{mackay1994bayesian}.

To estimate the posterior of the parameters ($\boldsymbol{\phi}, \sigma^2, \boldsymbol{w}$) given observations $\{(y_i, \boldsymbol{x_i})\}$ and hyperparameters ($\mu_\phi, \lambda_\phi, \alpha_{\sigma^2}, \beta_{\sigma^2}, \alpha_{w}, \beta_{w}$), we use mean-field variational inference \cite{attias1999inferring}. Initial values of the variables are drawn from the prior. The estimates of the relevance values are calculated by using the mean of the posterior distribution of $\boldsymbol{\phi}$.\footnote{For keywords that the user had given explicit feedback to, we adjusted the relevance value to be the mean of a given feedback and estimated relevance. The reason for this was that in a pilot study the users sometimes complained that the keywords did not go where the user dragged them. This "control problem" was also discussed in \cite{Kangasraasio2015} and our approach is a simplified way to address it.}

Variational inference on a linear Gaussian model with individual noise levels for observations was first introduced in \cite{tipping2003variational}. A similar model has been used successfully for outlier detection in robotics \cite{Ting07automaticoutlier}. Our model differs from it by allowing the user to correct the inferences, and also by estimating $\sigma^2$ with variational inference instead of using a point estimate. Taking full distributions into account is important because only a very small amount of user feedback is available for fitting the model.

In this particular scenario, we made a model for the relevance of the various keywords that appear in the recommended documents. The feature vectors of the keywords were constructed dynamically based on the TF-IDF scores of the keywords in the top 400 documents. The feature vectors were normalized to unit length (L2 norm). User relevance feedback was in the range $[0, 1]$, where larger values indicate higher relevance. The documents were ranked based on the estimated relevance of the most relevant keywords (more details can be found in \cite{Glowacka2013}).

\begin{figure*}[!hbt]
\centering
\includegraphics[width=0.97\textwidth]{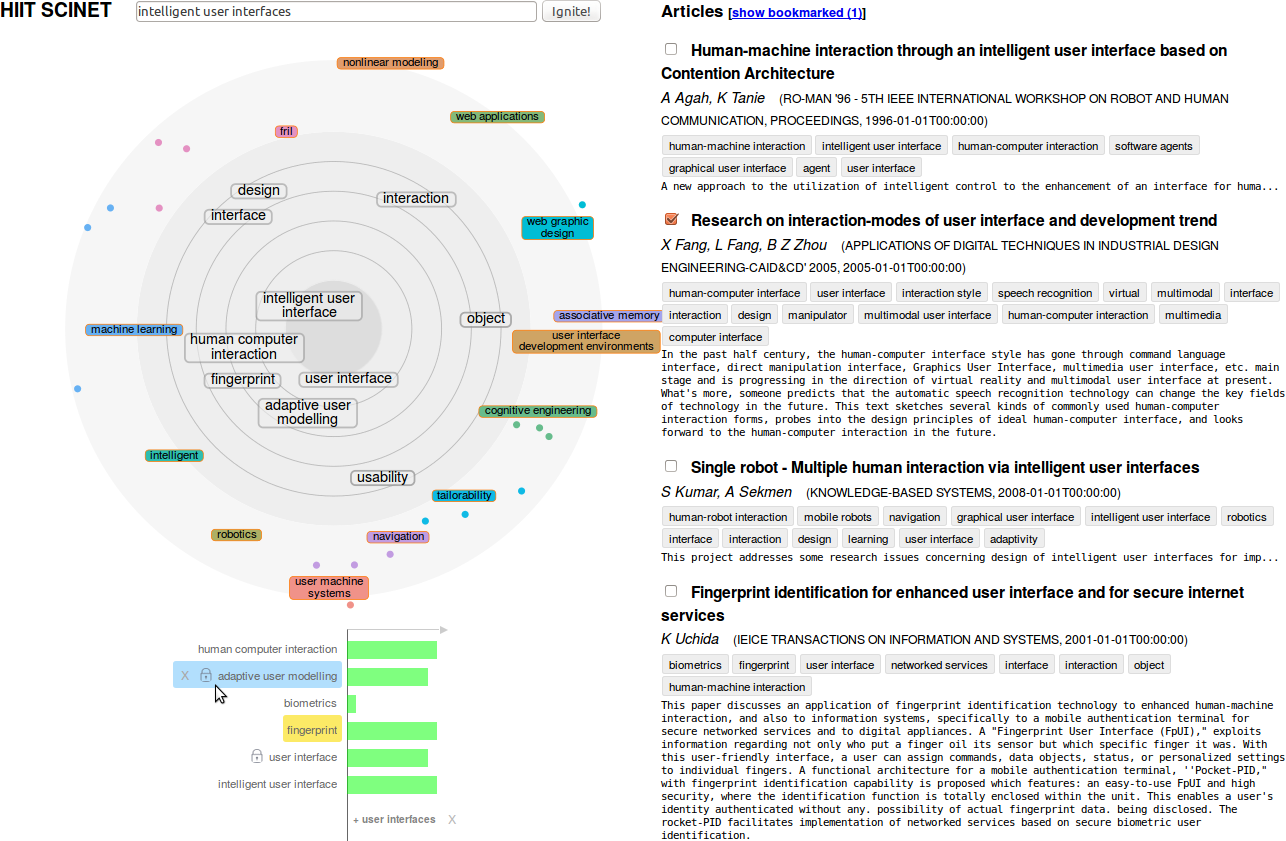}
\caption{The system's search interface. At the top of the screen there is  a keyword search bar.  Below it, a visualization of the current user model is shown using a radar metaphor: relevant keywords are closer to the center and less relevant ones are further away from the center. At the very edge of the radar there are keyword suggestions. At the bottom of the screen is the proposed timeline interface, showing the feedback the user has given so far in the current search session, as well as lists of keywords used in past sessions. On the right hand side of the radar visualisation there is a list of the most relevant documents. The abstract of each document can be expanded by clicking on it. The user can also bookmark articles to be able to access them at a later time by opening the bookmark list (blue link at the top).}
\label{fig:interface}
\end{figure*}

\section{User Interface}

The user interface is presented in Figure \ref{fig:interface}. It is largely similar to the interface presented in \cite{Kangasraasio2015}. The search results (a list of 10 most relevant documents recommended to the user) are displayed on the right side of the screen. On the left side, the user intent model is presented as a radar visualisation. The user can adjust the model by moving keywords to new locations on the radar (i.e. provide  relevance feedback to the keywords). The timeline interface is situated under the radar to display the user's history of relevance feedback.

The search starts with the user typing in an initial query in the search bar. This initial query is then transformed into a corresponding set of relevance feedback,\footnote{These "pseudo feedbacks" were generated by finding the most common keywords that appear in the documents that were retrieved based on the initial query alone. Keywords that were at least half as common as the most common keyword were selected. The feedback values were in proportion to the frequency of appearance, so that the most common keyword received feedback $1.0$. This relevance feedback was used to initialize the user model.} which is added to the timeline and used to fit the initial user model. The top 10 relevant keywords then appear in the center of the radar visualisation, and the list of recommended documents is presented on the right hand side of the screen.

\begin{figure}[!hbt]
\centering
\includegraphics[width=0.9\columnwidth]{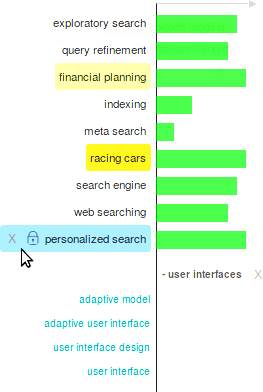}
\caption{The timeline interface visualizes past feedback (recent feedback on top) and provides the user with ways to interact with it. The interface for deleting and marking feedback as accurate is hidden unless the user hovers the mouse on top of the keyword (blue background with mouse). Feedback most likely in need of revision is highlighted with yellow background, which is darker for feedback that is more of an outlier. Highlighted feedback has an explanatory tooltip: "Is this feedback still accurate?". The user can click the lock icon to indicate that she is sure that it is accurate. The user can also adjust the relevance of a given keyword using the green bar. The feedback can be removed by clicking the X icon. The user can give feedback to keywords used in previous sessions by clicking on the area right of the keyword (blue keywords at the bottom). Hovering the mouse above the area shows a blue bar visualizing the feedback value.}
\label{fig:timeline}
\end{figure}

While the user is performing the search, the timeline visualisation (Figure~\ref{fig:timeline}) displays the keywords used so far in chronological order. The most recent feedback given by the user appears at the top of the timeline. Each feedback on the timeline has a green bar on the right hand side of the timeline to indicate the relevance of its keyword. The longer the bar, the higher the relevance the user has given to the keyword.

The motivation for the design of the timeline was to provide the user with a quick overview of the feedback that she has given in the current search session. In longer sessions, it is likely that the user will not remember the details of the feedback she has given earlier. The timeline helps the user to reflect on their earlier feedback and evaluate the feedback given so far as a whole. For example, when concept drift happens, some of the earlier feedback may no longer be valid with respect to the user's new interest.

The timeline interface was also designed to allow the search engine to notify the user if it suspects that some of the feedback is no longer accurate. This could be, for example, because the user has made an error in giving  feedback or because of concept drift. Feedback with low estimated accuracy is made salient to the user by yellow backgrounds.\footnote{Feedback is highlighted with increasing intensities when the estimated accuracy $w_i$ is below the threshold values $0.65$ (light yellow), $0.55$ (medium yellow) or $0.45$ (dark yellow). These values were tuned by hand.} Highlighting is expected to help the user to find feedback in need of revision more efficiently. The user can adjust the relevance value of a feedback by clicking on the corresponding position on the bar. She can also indicate that a feedback is accurate by clicking the lock icon, or delete a feedback by clicking the X-icon (removes the effect of the feedback from the model). The option to react to both true and false highlights (respectively by adjusting keyword relevance and marking feedback as accurate) was motivated by the results of the simulation experiment.

Keywords from previous search sessions are added as expandable lists at the bottom of the timeline. These lists can be removed by clicking the X-icon. The motivation for this feature was to provide the user with a convenient way to re-use keywords she interacted with in previous sessions.

\section{Simulation Experiment}

To study the performance of the user model, we conducted an experiment with a simulated user. As a dataset we used the 20 Newsgroups dataset \cite{20NGweb} containing 2000 newsgroup messages, 100 from each of 20 newsgroups. L2-normalized TF-IDF feature vectors of length 539 were generated for the posts. Terms with document frequency over $0.2$ or under $0.04$ were thresholded to remove too rare and too common terms.

In each repeated experiment the simulated user selected at random one of the 20 newsgroups as the search target. The user then initialized the query by indicating two positive examples from this group at random. The user model replied with a list of 50 most relevant documents and, depending on the scenario, one highlighted past feedback the user should re-evaluate. The user then replied by giving noisy feedback to one item in the list of 50 and, depending on the scenario, by possibly revising the highlighted feedback. This cycle was repeated for 100 steps. After each step, the F1-score of the list of 50 items was calculated (representing the quality of found items).

\begin{figure*}[!hbt]
\centering
\includegraphics[width=0.97\textwidth]{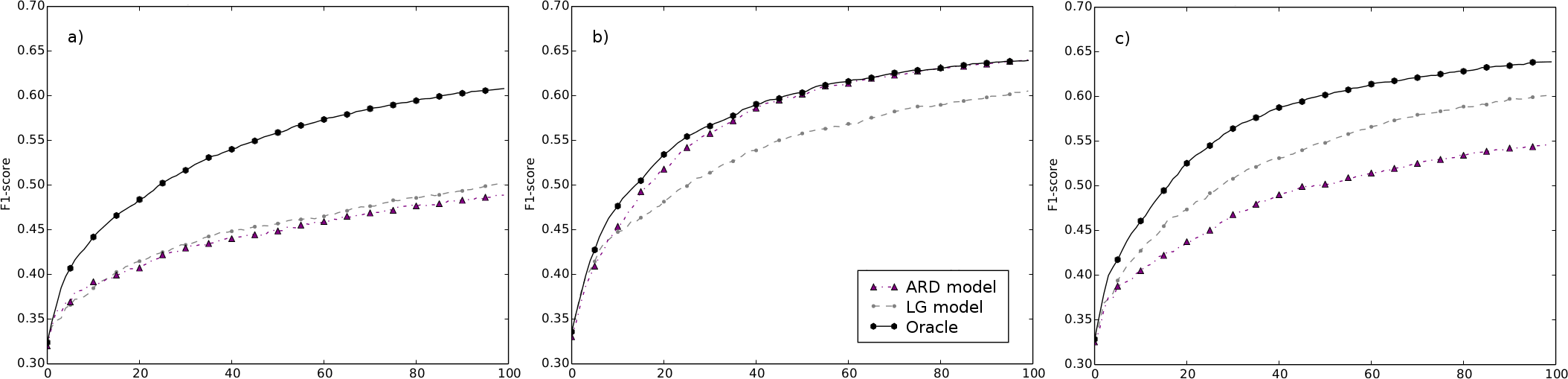}
\caption{Simulated F1-scores during 100 feedback iterations, averaged over 200 search sessions, in three of the four different scenarios. a) No items highlighted to the user. b) One item highlighted to the user at each step, user revises true positives and indicates false positives. c) One item highlighted to the user at each step, user revises true positives but does not indicate false positives. The results from the scenario where the user indicates false positives but does not revise true positives was similar to a) and was left out to save space.}
\label{resultfigure1}
\end{figure*}

The user's noisy feedback was generated as follows. From the list of items, $70\%$ of the time the user selected a positive example and gave it  relevance feedback with value $1.0$. The user selected a negative example $10\%$ of the time, and gave it relevance feedback $0.0$. $20\%$ of the time the user selected a random item from the list and gave it relevance feedback $1.0$ with $87.5\%$ probability or $0.0$ with $12.5\%$ probability. The proportions of negative and positive feedback reflect our past experience with user behavior with similar systems. The proportion of noise was chosen to be small,\footnote{Assuming half of the items in the list were relevant, the average proportion of false positive feedback was $1.25\%$ and false negative feedback was $8.75\%$} but sufficient to demonstrate the effect of the new user model.

The experiment was repeated in four different scenarios. In Scenario A, no items were highlighted to the user, and thus the user made no revisions to given feedback. In Scenario B, the user revised the highlighted feedback if it did not have the correct relevance value (i.e. revised true positives) and indicated that the feedback was was accurate if it already had the correct relevance value (i.e. indicated false positives). In Scenario C, the user only revised true positives, and in Scenario D the user only indicated false positives.

We compared the performance of the ARD model to a baseline and an oracle. The baseline was a Linear Gaussian model that was otherwise similar to the ARD model, except that all feedback was assumed to be equally accurate (i.e. $w_i = 1.0$). We will call this the \textit{LG model}. The \textit{Oracle} knew which feedback was relevant which was not and only used the relevant observations in fitting the model,  otherwise being similar to the LG model.

The ARD model chose the feedback to be highlighted by selecting the feedback having the lowest $w_i$ value. Draws were resolved randomly. The LG model sampled the highlighted feedback uniformly and the Oracle highlighted feedback based on the true relevance values.

The prior parameters of all the algorithms were hand tuned over a small number of iterations to avoid over-fitting. The used parameters were $\mu_\phi=0.0, \lambda_\phi=0.1, \alpha_{\sigma^2}=2.5, \beta_{\sigma^2}=0.5, \alpha_{w}=0.7$ and $\beta_{w}=1.0$. All the models were fit using  variational approximation. The convergence criterion was set to be an absolute change of less than $0.1$ in the unnormalized KL-divergence. The initial state was drawn from the prior.

The retrieval performance is shown in Figure~\ref{resultfigure1}. We observe that the ARD model performed similarly to the LG model if no corrections were made to the historical feedback (Scenario A). If the user reacted to both true and false highlights (Scenario B), the performance of the ARD model approached that of the Oracle. The LG model did not improve as much in this case, as the method for selecting the highlighted items was random. If the user made corrections to only the true positive highlighted items, the ARD model improved the retrieval performance from Scenario A, but the performance was not as good as in Scenario B. Surprisingly, the LG model performed better than the ARD model in this scenario. The reason for this was that if the ARD model made a mistake in identifying an outlier, this was not corrected by the user in this scenario, and thus the model always highlighted this particular item instead of trying some other items. In comparison, the LG model would eventually find the correct items to highlight through random sampling. If the user made corrections to only the false positive highlighted items (Scenario D), the improvements were small.

We also measured the average runtimes of the models per step in wall clock time. Averaged over all scenarios, after 10 steps ARD had average runtime of $0.6$ s, whereas LG had $0.4$ s. After 100 steps the average runtimes were $1.4$ s for ARD and $0.8$ s for LG. Simulations were made in a computing cluster equipped with 2.6 GHz processors.

Overall, we observed that the ARD model provided improvements over the simple baseline without increasing the runtime considerably. We also observed that the user should be able to react to both true and false highlights for the best performance when using the ARD model.

In this experiment the user model directly estimated the relevance of the newsgroup posts, instead of the keywords appearing in the posts. This was done to simplify the situation. However, the general results should apply to the more complex case as well, where first estimate the relevance of keywords and then order the documents based on the most relevant keywords.

\section{User study methodology}

We ran a user study to understand the effects of the timeline interface, combined with the new user modeling, on exploratory search tasks. We compared the proposed interface and user model to a baseline with a simpler user model and interface.

In the baseline, the timeline visualization was hidden from the user and the LG model was used for predicting the relevance of keywords. Thus the differences in user behavior are generally attributable to either the new user model or the new interface.

Eighteen participants (three female), aged 20 -- 30, took part in the user study. All of the participants were university students. Each participant was compensated with a movie ticket worth approximately 10 EUR.

Each participant performed two tasks -- one with each interface. The order of tasks and  interfaces was balanced as was the pairing of interfaces with tasks. Before performing the two tasks, the participants were shown a video tutorial on how to use the two interfaces. This was followed by a 30-minute practice session to allow the users to familiarize themselves with the systems at their own pace. In the practice session the participants were instructed to perform a free search related to their own study interests. The timeline interface was used in the practice session as it covers all the features present in both interfaces.

In the main tasks, the user was instructed to write a short draft of an essay on a given topic. The task descriptions following the template: "Many types of X exist in the field of Y. Try to find up to five types of X used in Y. Give some concrete examples of different X. Write your answer as an essay draft and bookmark at least 10 relevant articles, that you could use as a reference in writing the article." The two topics were "algorithms used in robotics" and "examples of information retrieval systems". Before performing each task, participants rated their familiarity with the topic on a 5-point Likert scale. All users reported familiarity between 1 and 4 (average $1.9$ for task 1, $2.1$ for task 2). The duration of each task was 20 minutes with a short break between the tasks.

After each task, the participants completed the SUS \cite{brooke1996sus} questionnaire with 10 questions (Table \ref{tab:sus_scores}) and a modified version of the ResQue questionnaire \cite{pu2011user} with 19 questions (Table~\ref{tab:resque_scores}). The ResQue questions were the same as the ones used in \cite{Kangasraasio2015} with four additional questions (No. 10, 11, 13 and 15 in Table~\ref{tab:resque_scores}). The aim of the additional questions was to learn how useful the participants found the timeline visualisation for conducting exploratory search.

During the experiments we logged the keywords seen and manipulated by the users at each iteration, the documents presented to the users, as well as the documents bookmarked by the users. After both tasks were completed, we conducted a semi-structured interview with the participants.

The parameters of the user model were hand tuned over a small number of iterations to avoid over-fitting. The used parameters were: $\mu_\phi=0.0, \lambda_\phi=0.1, \alpha_{\sigma^2}=2.0, \beta_{\sigma^2}=0.1, \alpha_{w}=1.0$ and $\beta_{w}=1.0$. The algorithm was limited to 10 iterations of the variational fitting to guarantee fast on-line performance. Based on initial tests the algorithm often converged before the limit.

\section{User Study Results}
Below, we report on the analysis of the user study results. Two participants were excluded from the analysis as they were not able to complete one of the tasks successfully.\footnote{Users were excluded from analysis if two independent experts rated their task performance as 1 out of 5 in one task}

\subsection{Questionnaire Results}
The SUS scores were similar for both interfaces. The average score was $68$ for the baseline and $72$ for the new interface ($p = 0.7$)\footnote{The reported p-values were calculated with the paired two-sided Wilcoxon signed-rank test and rounded up. Each p-value was calculated independently and was reported as such for completeness.}. This indicates that the usability of the system did not suffer from the added functionality.

The new interface got better ResQue scores. The average score was $50$ for the baseline and $55$ for the new interface ($p = 0.04$). Per-question scores and p-values are shown in Tables~\ref{tab:sus_scores}~and~\ref{tab:resque_scores}. User ratings indicate that the new user model generates better and more diverse results (ResQue 2, 3, 16). It was easier for users to notice and correct mistakes in previous feedbacks using the timeline interface (ResQue 10, 13). The timeline interface also seemed to make it easier for the users to modify their query (ResQue 9, 11). The users also seemed to prefer the new interface layout compared to the baseline (ResQue 4, 6).

\begin{table}[!h]
  \centering
  \begin{tabular}{| >{\centering}p{0.4cm} | >{\centering}p{0.4cm} | >{\centering}p{0.4cm} | p{5.8cm} |}
    \hline
    \tabhead{N} &
    \tabhead{B} &
    \tabhead{p} &
    \tabhead{Question} \\
    \hline
\small{\textbf{3.8}} & \small{\textbf{3.8}} & \small{0.9} & \small{1: I think that I would like to use this system frequently}\\
\small{2.6} & \small{\textbf{2.3}} & \small{1.0} & \small{2: I found the system unnecessarily complex}\\
\small{\textbf{3.9}} & \small{\textbf{3.9}} & \small{1.0} & \small{3: I thought the system was easy to use}\\
\small{\textbf{2.0}} & \small{\textbf{2.0}} & \small{1.0} & \small{4: I think that I would need the support of a technical person to be able to use this system}\\
\small{\textbf{3.6}} & \small{\textbf{3.6}} & \small{0.8} & \small{5: I found the various functions in this system were well integrated}\\
\small{\textbf{2.2}} & \small{2.9} & \small{0.2} & \small{6: I thought there was too much inconsistency in this system}\\
\small{4.3} & \small{\textbf{4.4}} & \small{0.4} & \small{7: I would imagine that most people would learn to use this system very quickly}\\
\small{2.1} & \small{\textbf{2.0}} & \small{0.9} & \small{8: I found the system very cumbersome to use}\\
\small{3.9} & \small{\textbf{4.0}} & \small{0.6} & \small{9: I felt very confident using the system}\\
\small{\textbf{1.8}} & \small{1.9} & \small{0.8} & \small{10: I needed to learn a lot of things before I could get going with this system}\\
    \hline
  \end{tabular}
  \caption{SUS score question averages for the new interface (N) and the baseline (B) system with p-values. Questions were scored on a 5-point Likert scale from 1 (disagree) to 5 (agree). The better value in each row is in boldface; higher is better for odd numbered questions and lower is better for even numbered questions.}
  \label{tab:sus_scores}
\end{table}

\begin{table}[!h]
  \centering
  \begin{tabular}{| >{\centering}p{0.4cm} | >{\centering}p{0.4cm} | >{\centering}p{0.4cm} | p{5.8cm} |}
    \hline
    \tabhead{N} &
    \tabhead{B} &
    \tabhead{p} &
    \tabhead{Question}\\
    \hline
\small{\textbf{4.1}} & \small{3.9} & \small{0.3} & \small{1: The items recommended to me matched what I was searching for}\\
\small{\textbf{4.6}} & \small{4.1} & \small{0.02} & \small{2: The recommender system helped me discover new items}\\
\small{\textbf{4.0}} & \small{3.5} & \small{0.05} & \small{3: The items recommended to me are diverse}\\
\small{\textbf{3.8}} & \small{3.4} & \small{0.08} & \small{4: The layout of the recommender interface is adequate}\\
\small{\textbf{3.6}} & \small{3.2} & \small{0.2} & \small{5: The recommender explains why the items are recommended to me}\\
\small{\textbf{3.8}} & \small{3.4} & \small{0.2} & \small{6: The information provided for the recommended items is sufficient}\\
\small{\textbf{3.6}} & \small{3.4} & \small{1.0} & \small{7: I found it easy to tell the system what I want / don't want to find}\\
\small{4.1} & \small{\textbf{4.3}} & \small{0.5} & \small{8: I became familiar with the recommender system very quickly}\\
\small{\textbf{4.2}} & \small{3.8} & \small{0.2} & \small{9: I found it easy to modify my search query in the recommender}\\
\small{\textbf{3.6}} & \small{3.1} & \small{0.06} & \small{10: I found it easy to notice if some of my query modifications were not correct any more}\\
\small{\textbf{3.9}} & \small{3.6} & \small{0.3} & \small{11: I found it easy to find suitable ways to modify my query}\\
\small{\textbf{3.9}} & \small{3.5} & \small{0.05} & \small{12: I understood why the items were recommended to me}\\
\small{\textbf{3.5}} & \small{2.9} & \small{0.09} & \small{13: I found it easy to notice if I had made a mistake in modifying my query}\\
\small{\textbf{3.9}} & \small{3.6} & \small{0.05} & \small{14: Using the recommender to find what I like is easy}\\
\small{\textbf{3.5}} & \small{3.2} & \small{0.3} & \small{15: I found it easy to re-find items I had been recommended before}\\
\small{\textbf{4.3}} & \small{4.0} & \small{0.2} & \small{16: The recommender gave me good suggestions}\\
\small{\textbf{4.0}} & \small{3.8} & \small{0.5} & \small{17: Overall, I am satisfied with the recommender}\\
\small{\textbf{4.3}} & \small{4.0} & \small{0.3} & \small{18: The recommender can be trusted}\\
\small{\textbf{4.1}} & \small{3.9} & \small{0.5} & \small{19: I would use this recommender again, given the opportunity}\\
    \hline
  \end{tabular}
  \caption{ResQue score question averages for improved (I) and baseline (B) system with p-values. Questions were scored on a 5-point Likert scale from 1 (disagree) to 5 (agree). The better value in each is in boldface; higher is better.}
  \label{tab:resque_scores}
\end{table}

\subsection{Log Data Analysis}
We logged the actions of the users during the experiment. The users performed on average $5.6$ keyword queries per task with the baseline and $3.8$ with the new interface ($p = 0.2$). The number of keyword-related interactions (giving feedback to a keyword, removing or marking feedback as accurate) was larger with the new interface. Users did on average $5.5$ keyword interactions per task with the baseline and $10.8$ with the new interface ($p = 0.001$). The interactions with the new interface consisted of on average $6.8$ keyword feedback on the radar ($p = 0.09$ compared to baseline), $1.2$ keyword feedback on the timeline, $1.9$ keyword deletions from the timeline, $0.9$ feedback marked as accurate and $0.1$ feedback given to archived keywords from past search sessions. These results indicate that  users interacted more frequently with the new system. This was not entirely due only to the fact that they had more interaction options, as on average users also performed more interactions with the radar when using the new interface. Users also seemed to write fewer keyword queries, likely resulting from the increased interaction options.

Various proxies for  users' engagement and quality of the retrieved results were also monitored.  Users expanded to view on average $15$ articles' abstracts with the baseline and $17$ with the new interface ($p = 0.5$). The average number of viewed unique articles per task was similar for both interfaces: $61$ for baseline and $63$ for new interface ($p = 0.4$). The average numbers of viewed unique keywords per task were $43$ central and $197$ peripheral for the baseline, and $41$ central and $233$ peripheral for the new interface (local: $p = 0.8$, peripheral: $p = 0.1$). It appears that the timeline interface provided more diverse keyword suggestions to the user.

\subsection{Expert Evaluations}
Task performance was assessed in a blind manner by two independent experts based on the written answers and bookmarked articles. The ratings were done on 5-point Likert scale from 1 (bad) to 5 (good). The average task performance was $3.6$ with the baseline and $3.5$ with the new interface ($p = 0.6$). This indicates that there was no significant difference in the task performance between the two systems. Inter-rater reliability\footnote{Inter-rater reliability was calculated with Spearman's rho and rounded down.} was $0.6$ for both tasks.

We asked an expert to evaluate the keywords shown to the users in the center of the radar. The keywords were divided into three categories: \textit{general}, containing keywords that are generally relevant to the topic; \textit{specific}, containing keywords that are specifically relevant to the topic; and \textit{irrelevant}, containing keywords that are not relevant to the topic. The proportions of general, specific and irrelevant keywords shown on average to the user were $41\%$, $47\%$, $11\%$ with the baseline and $48\%$, $47\%$, $5\%$ with the new interface ($p = 0.6, 1.0, 0.2$). This indicates that the keywords shown to the user were slightly more relevant with the new system compared to the baseline.

We also asked an expert to evaluate the articles shown to the users. The articles were divided into three categories: \textit{obvious}, containing common articles related to the topic; \textit{novel}, containing articles that are less common but relevant to the topic; and \textit{irrelevant}, containing articles that are not relevant to the topic. The proportions of obvious, novel and irrelevant articles shown on average to the user were $7\%$, $80\%$, $13\%$ with the baseline and $6\%$, $81\%$, $13\%$ with the new interface ($p = 0.3, 0.8, 1.0$). This indicates that the quality of the articles shown to the user were approximately the same between the systems.

\subsection{User Interview Analysis}
After the main tasks, we conducted a semi-structured interview with each user. Almost all the users reported that they preferred the timeline interface to the baseline. The most often mentioned benefits of the timeline interface were:
\begin{enumerate}
    \item Helped users to track and compare relevance of keywords they had interacted with,
    \item Gave the user subjectively more control over the system, as the users felt that the relevance bars in the timeline make it easier and more accurate to set and modify the relevance of keywords,
    \item Enabled users to re-use keywords from past search sessions.
\end{enumerate}

The delete function was used mostly as we expected. For example, users reported to have removed feedbacks which were no longer valid when switching to another sub-topic, or when they wanted to remove the effect of a particular feedback. According to the interviews, many users did not appreciate the function of "marking feedback as accurate", and several users reported having not used this function at all. However, a few users used it in a creative way, for example, "locking" the feedback to the core keywords related to the topic while trying to explore  different sub-topics. The highlighting of feedback was reported to attract attention, and some users tried to respond to each of them, although some users felt that sometimes too many keywords were highlighted. The users also made the following suggestions as features they would like to add to the system:
\begin{enumerate}
    \item Ability to perform multiple queries simultaneously instead of being restricted to one active session at a time,
    \item Ability to give feedback to multiple keywords at once rather than just one at a time,
    \item Availability to go back to a specific previous state,
    \item Ability to see more search results for the same query if needed.
\end{enumerate}

\newpage
\section{Discussion and Conclusions}

We introduced a new user model that is able to take into account concept drift and user errors in relevance feedback as well as a timeline interface that allows the user to interact with the new model and  view her relevance feedback history.

In a simulation experiment we showed that the new user model is able to improve the search results over a baseline when the user responds to the highlights made by the model. However, we also noticed that for best performance, the user needs to be able to respond to both true and false positive highlights.

Additionally, we conducted a user study, where we measured multiple variables related to user performance, satisfaction and actions with the system.
Users reported that the new system made it easier for them to notice and correct mistakes in their feedback, and that they felt the it was easier for them to modify their query with the new interface. Users also seemed to prefer the new interface layout to the baseline.

In general, users felt that the new system helped them to discover new items, the quality of the results was better and that the results were more diverse with the new system. However, the number of unique articles and keywords shown to the user were mostly the same with both systems. The expert-rated quality of the keywords and articles shown to the users was also similar between the two systems. One explanation for this is that we did not evaluate the quality of the results per query but with respect to the general relatedness to the topic of a given search task. It may be that users were occasionally exploring areas not directly related to the topic, and thus the subjective result quality could well be higher than the results rated by the expert. 

Users interacted more frequently with the new interface, and this was not only because of the new interaction options related to the presence of the timeline -- users also gave more feedback using the radar interface in the new system. Additionally,  users issued fewer  keyword queries with the new interface, indicating that the new interface options made it easier for them to modify the query in other ways. These changes did not seem to have any large effect on the task performance, which was similar with both systems. Also, there was no indication that the usability of the system decreased because of the added functionality.

In post-experiment interviews, users reported that the new interface allowed them to easily track and compare the feedback they had given, enabled them to re-use keywords from past sessions and that the new interface gave them subjectively more control over the system. The usage of the added functionality was mostly as expected, although  users also found novel ways to use the functions that we had not thought of before the experiment.

To the best of our knowledge, this is the first search system that both models the accuracy of individual user feedback in a search setting and allows the user to directly interact with this model.

\newpage
\section{Acknowledgments}

This work has been partly supported by the Academy of Finland (Finnish Centre of Excellence in Computational Inference Research COIN) and TEKES (Re:Know). The research leading to this results has received funding from the European Union Seventh Framework Programme (FP7/2007-2013) under grant agreement no 611570. We acknowledge the computational resources provided by the Aalto Science-IT project.

\bibliographystyle{acm-sigchi}
\bibliography{paper}
\end{document}